\documentclass[preprint,12pt]{elsarticle}
\usepackage{graphicx}
\usepackage{epstopdf}
\usepackage{subfigure}
\usepackage{url}
\usepackage{natbib}

\journal{}
\usepackage{float}
\usepackage{amssymb}
\usepackage{lineno,hyperref}
\usepackage{color}
\modulolinenumbers[5]
\usepackage{lineno}

\graphicspath{{images/}}
\newcommand{\vect}[1]{\vec{#1}}
\newcommand{\vecm}[1]{\mathbf{#1}}

\modulolinenumbers[1]
\begin{document}

\begin{frontmatter}
\title{A crosstalk and non-uniformity correction method for the space-borne Compton polarimeter
POLAR}
\author[h4,h1] {Hualin Xiao\corref{xiaohlcor}}
\ead{hualin.xiao@psi.ch}
\author[h4]{Wojtek Hajdas}
\author[h1]{Bobing Wu}
\author[h2]{Nicolas Produit}
\author[h1]{Tianwei Bao}
\author[h5]{Tadeusz Batsch}
\author[h3]{Franck Cadoux}
\author[h1]{Junying Chai}
\author[h1]{Yongwei Dong}
\author[h1]{Minnan Kong}
\author[h1]{Siwei Kong}
\author[h5]{Dominik K. Rybka}
\author[h3]{Catherine Leluc}
\author[h1]{Lu Li}
\author[h1]{Jiangtao Liu}
\author[h1]{Xin Liu}
\author[h4]{Radoslaw Marcinkowski}
\author[h3]{Mercedes Paniccia}
\author[h3]{Martin Pohl}
\author[h3]{Divic Rapin}
\author[h1]{Haoli Shi}
\author[h1]{Liming Song}
\author[h1]{Jianchao Sun}
\author[h5]{Jacek Szabelski}
\author[h1]{Ruijie Wang}
\author[h1]{Xing Wen}
\author[h1]{Hanhui Xu}
\author[h1]{Laiyu Zhang}
\author[h1]{Li Zhang}
\author[h1]{Shuangnan Zhang}
\author[h1]{Xiaofeng Zhang}
\author[h1]{Yongjie Zhang}
\author[h5]{Ania Zwolinska}
\address[h4]{PSI, 5232 Villigen PSI, Switzerland}
\address[h1]{Key Laboratory of Particle Astrophysics, Institute of High Energy Physics, Beijing 100049, China}
\address[h2] {ISDC, University of Geneva,1290 Versoix, Switzerland}
\address[h3]{DPNC, University of Geneva, quai Ernest-Ansermet 24, 1205 Geneva, Switzerland}

\address[h5]{The Andrzej Soltan Institute for Nuclear Studies, 69 Hoza str., 00-681 Warsaw, Poland}

\cortext[xiaohlcor]{Corresponding author. Tel.: +41\ 766682608.}

\begin{abstract}

 In spite of extensive observations and numerous theoretical studies in the past decades several key questions related with 
 Gamma-Ray Bursts (GRB) emission mechanisms are still to be answered. Precise detection of the GRB polarization
 carried out by dedicated instruments can provide new data and be an ultimate tool to unveil their real nature.
 A novel space-borne Compton polarimeter POLAR onboard the Chinese space station TG2 is designed to measure linear
 polarization of gamma-rays arriving from GRB prompt emissions.
POLAR uses plastics scintillator bars (PS) as gamma-ray detectors and  multi-anode photomultipliers (MAPMTs) 
for readout of the scintillation light. Inherent properties of such detection systems are crosstalk and non-uniformity.
The crosstalk smears recorded energy over multiple channels making both non-uniformity corrections and energy calibration more difficult.
Rigorous extraction of polarization observables requires to take such effects properly into account.
We studied influence of the crosstalk on energy depositions during laboratory measurements with X-ray beams.
A relation between genuine and recorded energy was deduced using an introduced model of data analysis. It postulates that both the
crosstalk and non-uniformities can be described with a single matrix obtained in calibrations with mono-energetic X- and gamma-rays. 
Necessary corrections are introduced using matrix based equations allowing for proper evaluation of the measured GRB spectra. 
Validity of the method was established during dedicated experimental tests. 
The same approach can be also applied in space utilising POLAR internal calibration sources.   
The introduced model is general and with some adjustments well suitable for data analysis from other MAPMT-based instruments.
\end{abstract}


\end{frontmatter}

\section{Introduction}

Gamma-ray bursts (GRBs) are observed as short flashes of gamma-rays appearing randomly in the sky.
In a few seconds they release energy between $10^{42}$ J and $ 10^{48}$ J making them the most energetic explosions in the Universe. 
They are frequently associate with either a collapse of the massive stars or a violent merge of compact binaries \cite{grblasts}.
In spite of numerous observations and theoretical efforts in the past decades
many key questions such as GRB emission mechanisms or origin and structure
of their magnetic field are not answered yet (for recent reviews, see Refs. e.g. \cite{grblasts, zb3, review1}).
Direct measurements of the GRB polarization in the prompt emission phase should be able to shed light on the whole system and
constrain the energy emission mechanisms \cite{zb2,zb1}.

To achieve this goal, several dedicated instruments are currently under development \cite{grape,nicolas,pogolite,poet,gap}.
They are specially designed and optimized for polarization measurement with support from precise on-ground calibrations as well 
as rigorous performance modeling and verification.
POLAR is one of such new hard X-ray polarimeters aimed to study the GRB prompt emissions.
It utilizes Compton scattering to measure linear polarization of  gamma-rays in the energy range from 50 keV to 500 keV.
Plastic scintillator (PS) bars are chosen as gamma-ray detecting medium and $8 \times 8$ channels multi-anode 
photomultipliers (MAPMTs) serve for readout of the scintillation photons.

It is an inherent property of MAPMT-based detectors that a signal from a certain channel can, through crosstalk effects
induce signals in the neighboring channels.
In addition, the response non-uniformity of the MAPMT modifies initially deposited energy altering the signal amplitudes.
Together with the crosstalk effect that spreads the initial energy deposition, it makes the energy calibration and spectral unfolding 
more difficult.
Thus, the precise knowledge of both effects is necessary to properly extract polarization observables in the detected GRBs.
 
Based on laboratory calibration data we constructed a model describing both the crosstalk and response non-uniformities 
for all 64 channels of each POLAR module.
The relation between genuine and recorded energy deposition can be described by a single matrix.
In the following chapters we describe methods applied to determine matrix elements and present results 
from laboratory tests used for their verification.

\section{POLAR instrument}

\begin{figure*}[htb]
\begin{minipage}{0.49\linewidth}
\includegraphics[width=0.95\textwidth]{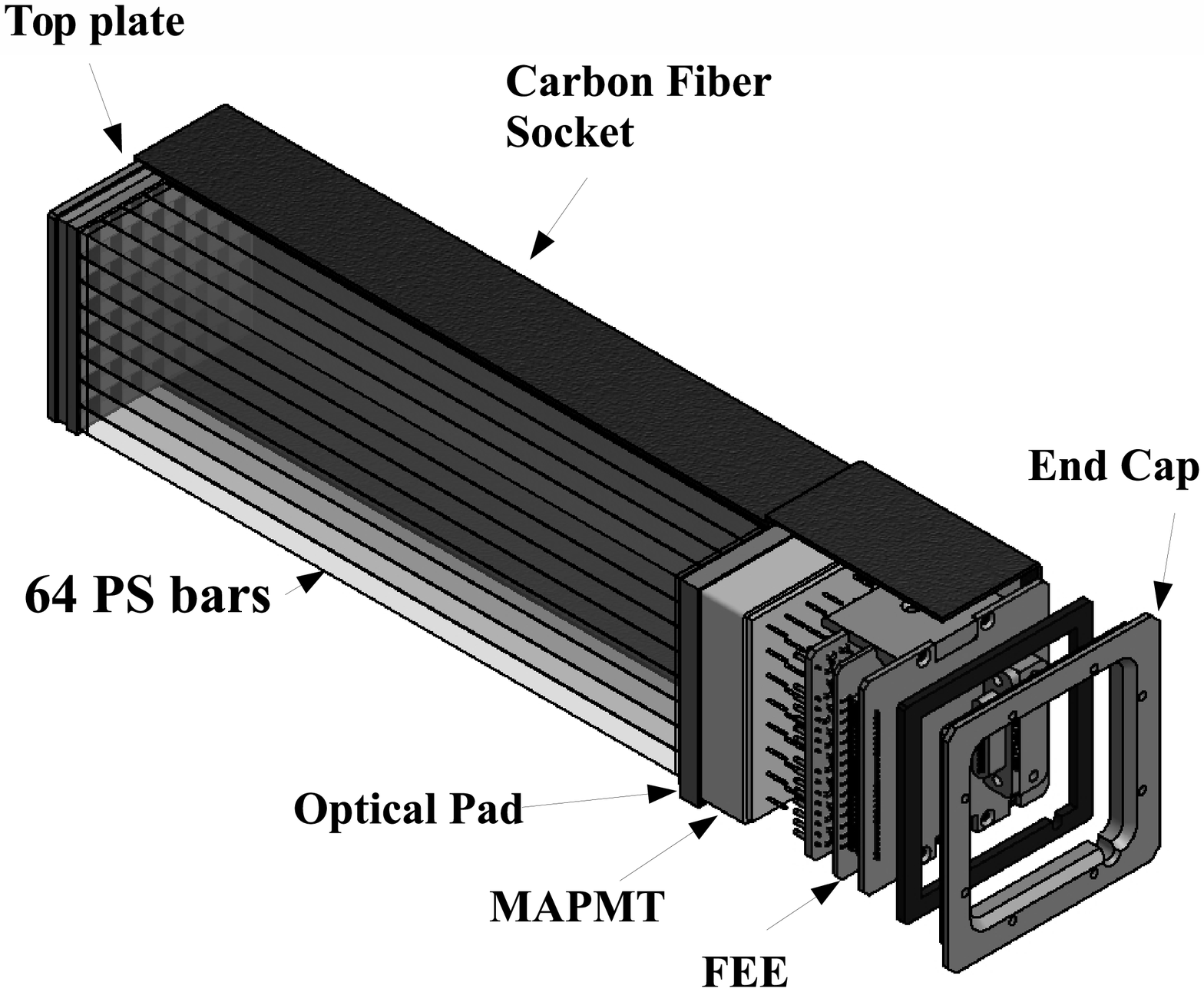}
\end{minipage}
\hspace{\fill}
\begin{minipage}{0.5\linewidth}
\includegraphics[width=0.95\textwidth]{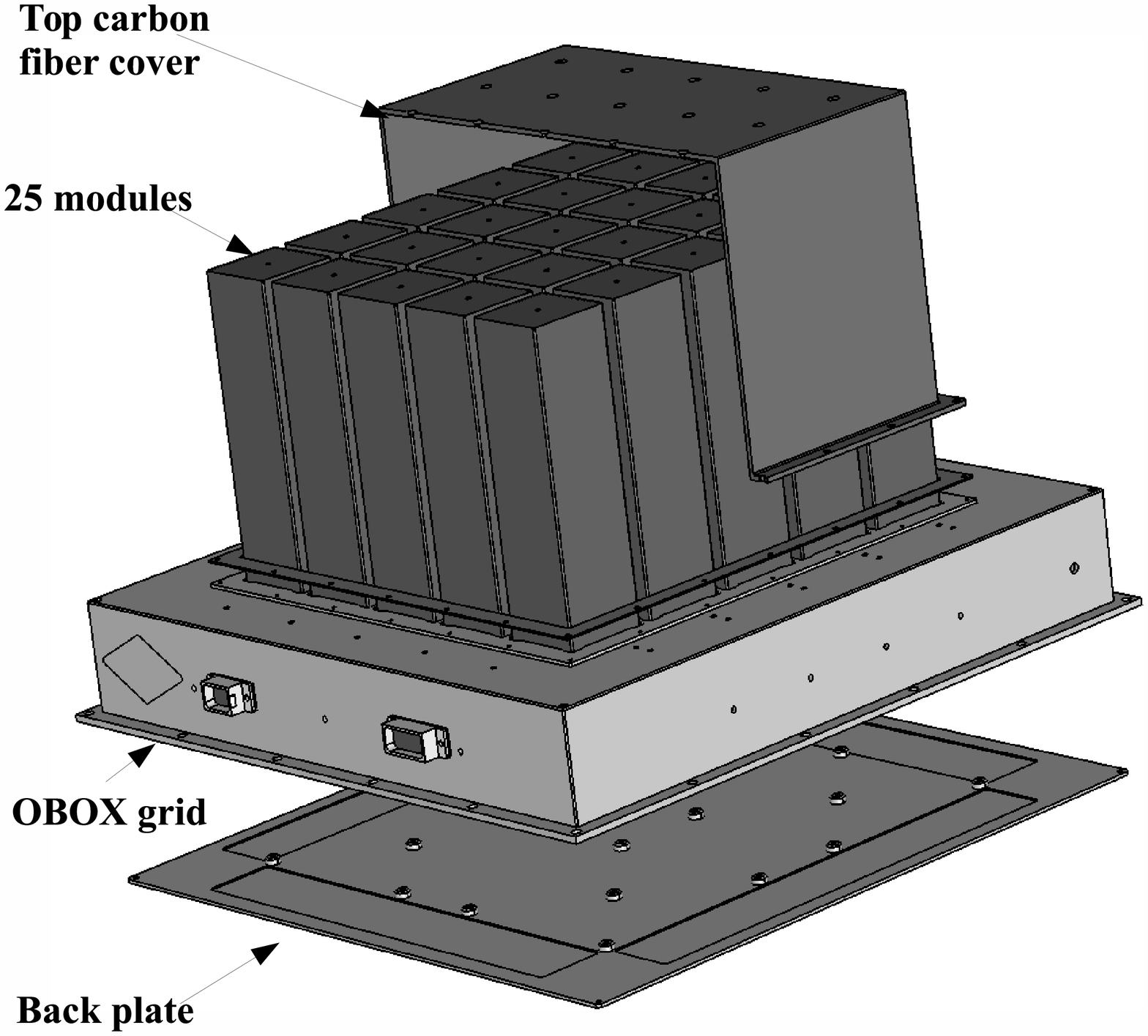}
\end{minipage}
\caption{Exploded view a POLAR detector module (left) and the full POLAR instrument (right).
Each module has 64 PS bars ($5.9 \times 5.9 \times 176\ \mathrm{mm}^3$ each) coupled to a 64 channel MAPMT
(Hamamatsu H8500)
and a front-end electronics.
The full instrument consists of 25 identical modules.}
\label{fig:polar}
\end{figure*}

The main goal of POLAR to measure linear photon polarization is realized using Compton scattering.
Polarized gamma-rays undergoing Compton process tend to scatter perpendicularly to their incident polarization vector according to the Klein-Nishina equation:
\begin{equation}
\frac{\mathrm{d}\sigma}{d\Omega}=
\frac{r_\mathrm{e}^2}{2}\left( \frac{E^{'}}{E} \right )^2
\left( \frac{E^{'}}{E}+\frac{E^{'}}{E}- 2 \sin{\theta}^2 \cos{\eta}^2 \right),
\end{equation}
where $r_\mathrm{e}$ is the classical radius of the electron, $E$ and $E^{'}$ are the energy of the incident photon and
the scattered photon, respectively, $\theta$ is the scattering angle between initial and final photon direction, 
and $\eta$ is the azimuthal scattering angle between the initial polarization vector and the direction of the scattered photon.

POLAR has both, a large effective detection area ($\sim$ 80 cm$^2$) and a wide field of view ($\sim$ 1/3 of full sky). 
They are needed for efficient and precise measurements of the  azimuthal distribution of gamma-rays undergoing Compton scattering in its scintillator bars.
After being hit by a gamma-ray the instrument records energy depositions in its 1600 channels.
POLAR uses an array of 40 (row) $\times$ 40 (column) plastic scintillator (PS) bars as gamma-ray detection target.
The azimuth angle of the scattered gamma-ray is determined from positions of two bars
with the highest energy depositions. 
Polarization degree as well as polarization angle of the detected GRB can be retrieved in the off-line reconstruction of all recorded gamma-ray events.

The scintillating material EJ-248 was chosen because of its fast response and high value of the softening temperature (90$^{\circ}{\rm C}$).
Each PS bar has dimensions of $5.9 \times 5.9 \times 176\ \mathrm{mm}^3$.
In order to reduce optical crosstalk all bars have narrower bottom-end cross-sections resembling a pyramid-like shape.
The procedure used for bar selection involved several strict criteria.
Firstly, all of them were carefully inspected to reject macroscopic defects.
Secondly, the dimension of each bar was precisely measured and only bars with dimension deviations smaller than 0.1 mm were accepted. 
Afterwards, the light output difference between the top and the bottom of each bar was measured with a dedicated setup. 
It consisted of a photomultiplier and an Aluminum-made bar holder lined with the Enhanced Specular Reflector (ESR) films. 
The test used a collimated Am-241 source placed at the bottom and the top of the bar. 
Only bars with the light output difference smaller than 10\% were accepted. 
Finally, to increase the light collection, all selected bars were wrapped in the 65 $\mu$m thick ESR film.

\begin{figure}[tb]
\centering
\includegraphics[width=0.5\textwidth]{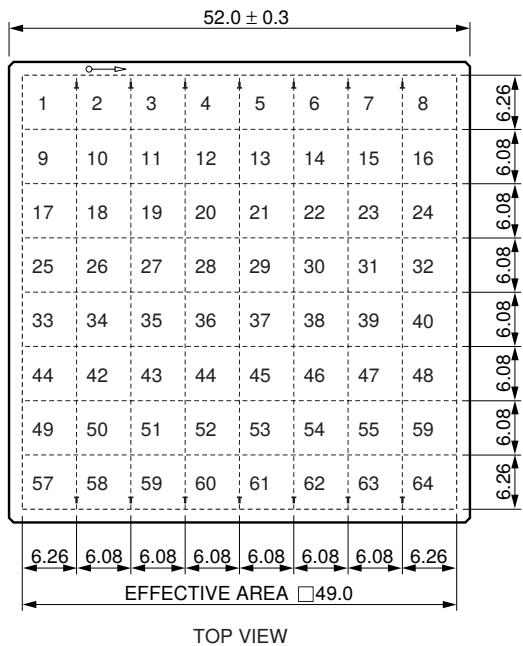}
\caption{Numbering convention of POLAR module channels adopted for data analysis from Hamamatsu H8500 MAPMT
\cite{h8500} (seen from the top of the glass window). 
The dimensions of the MAPMT channels are also shown.
}
\label{fig:h8500}
\end{figure}

The 1600 selected bars were assigned to 25 identical modules. 
Each module consists of 8$\times$8 PS bars, a soft optical coupling pad made of transparent silicon (Dow Corning DC93-500),
a 64 channel flat panel MAPMT (Hamamatsu H8500) and a front-end electronics (FEE). 
This structure was enclosed in a 1 mm thick carbon fiber sockets (see the left pannel of Fig.~\ref{fig:polar}). 
The top and the bottom of each PS were fixed and aligned with two plastic frames. 
They enhance resistance to vibrations and further reduce the optical crosstalk. 
The dimension of the optical coupling pad is $50 \times 50 \times 0.5$ $\mathrm{mm}^3$.
It can partially absorb vibrations and also protects the MAPMT glass window.
The channel numbering convention and MAPMT dimensioning are shown in Fig.~\ref{fig:h8500}.

The front-end electronics consists of three stacked Printed Circuit Boards (PCBs). The boards contain low voltage power supply circuits, a dedicated voltage divider for the MAPMT, a multi-channel ASIC (IDEAS VA64), an ADC, a FPGA and a temperature sensor. 
A special internal pulser circuit is also included to test the throughput and gain of each input channel. 
Moreover, at the inner edge of each corner-located module a low activity, point-like 22-Na radioactive source is installed.
Two annihilation photons (511 keV) from the source are used for the in-flight calibration done by applying the same data analysis techniques as outlined below.  
If the incoming signal has an amplitude above the threshold the ASIC sets a trigger flag that is sensed by the Central Task Processing Unit (CT).
If CT accepts the event, the front-end electronics digitises previously hold amplitudes from the 64 analog signals, 
packs them into the Science Data Packet and transmits to CT for further processing. 
In addition, the arriving time and the hit pattern of the trigger flags are also recorded by CT and form the Trigger Data Packet.
Dedicated part of CT also manages the low and high voltage power supply and the settings for all FEEs. 
It also handles communication with the spacelab.
All 25 modules, the CT, the power supplies and interfacing electronics are enclosed in a housing box as shown in the right panel of Fig.~\ref{fig:polar}.
The whole instrument is mounted on the outside pannel of of the Chinese TG-2 spacelab facing to the outer space.

Monte Carlo simulations showed that for each Compton scattered photon the line connecting  two PS bars with the maximum energy depositions is correlated with the outgoing azimuthal photon direction.
We define $\xi$ as the azimuthal angle between the line connecting these two bars and the x-axis of the POLAR detector. According to previous 
results  (see e.g. Refs. \cite {nicolas} and \cite{polarconstruction}), the distribution of $\xi$ from all detected GRB gamma-rays is called a modulation curve and follows characteristic pattern:  
\begin{equation}
f(\mu)=K\cdot\{1+\mu \cos[2(\xi-\xi_0)]\},
\end{equation}
where $K$ is the normalization factor, $\mu$ is the modulation factor and $\xi_0$ is the polarization angle.
The polarization degree $p$ is equal to $p=\mu/\mu_{100}$, where $\mu_{100}$ the instrument modulation factor  for 100\% polarized gamma-rays.
Realistic Monte Carlo simulations and calibrations with 100\% polarized x-rays are used to determine the value of $\mu_{100}$. 
The modulation factor $\mu$ and the polarization angle $\xi_0$ are obtained from the fit to the experimental data.
For detailed description of the reconstruction of both polarization observables see Refs.~\cite{nicolas,polarconstruction, silvio}.

\section{Crosstalk and response non-uniformity modeling}
Optical photos in PS bars are produced by ionizing particles  as result of their energy depositions.
The average number of optical photons $N_{\rm bar}$ collected
at the bottom of the bar is given by
\begin{equation}
N_{\rm bar}= s \cdot c \cdot E_{\rm dep},
\label{eq:lc}
\end{equation}
where
$E_{\rm dep}$ is the real energy deposition (i.e. visible energy deposition),
$s$ is the scintillation efficiency, i.e. the average number of optical photons produced per unit of deposited energy, and
$c$ is the photon collection efficiency, i.e. the fraction of the scintillation photons reaching the bottom of the bar.
The cumulative coefficient $c$ parametrizes effects of imperfections in  micro-cracks, reflectivity and attenuation of the optical light in the bars.
Knowing that all POLAR PS bars came from the same production batch and the in-flight temperature differences between them are minor,
it is reasonable to assume that they all have the same scintillation efficiency.
Hence, for each POLAR detector module, the number of photons reaching the bottom of its 64 PS bars 
$\vect{N}_{\rm bar} = (N_{\rm bar0}, N_{\rm bar1}, \cdots, N_{\rm bar63})^{\rm T}$
can be described as
\begin{equation}
\vect{N}_{\rm bar}= \vecm{B} \vect{E}_{\rm dep},
\label{eq:nbar}
\end{equation}
where the vector $\vect{E}_{\rm dep}= (E_{\rm dep,0}, E_{\rm dep,1}, \cdots, E_{\rm dep,63})^{\rm T} $
represents 64 real energy depositions in the bars,
and $\vecm{B}$ is a diagonal matrix expressed as
\begin{equation}
\vecm{B}= s \mathrm{Diag}(c_1 , c_2, \cdots, c_{64} ).
\end{equation}

Due to crosstalk effects  
any signal appearing in a certain channel can simultaneously induce signals in its neighboring channels. 
According to the H8500 MAPMT datasheet \cite{h8500}, 
a typical crosstalk factor between two neighboring channels  measured with an optical fiber of 1 mm diameter is of the order of 1\%.
Measurements in Ref.~\cite{mapmtRef2} also supports this value.
As this factor includes both optical and electrical contribution e.g. from stray or secondary electrons in the dynodes,  
one can conclude that the electrical part is smaller than 1\%. 
On the other hand, measurements with fully assembled POLAR modules showed that the total crosstalk between neighboring channels can be as high as 
10\% -- 20\%.  
This contains further contributions from stray photons in the optical pad and spreading of photons between PS bars. 
Additional tests showed that the electric crosstalk in the module readout system is negligible \cite{sunjc}.
As the overall electrical crosstalk is very small compared to the optical one, we only consider the later. 
In order to describe its strength, we introduce a $64 \times 64$ matrix
$\vecm{X}=\left(x_{ij}\right)$,
in which the matrix element $x_{ij}$
represents the fraction of photons in the $j$-th channel coming from the primary energy deposition in the $i$-th channel.
Obviously, we have $0 \leq x_{ij}\leq 1$ and $\sum\limits_{j=0}^{63} x_{ij} = 1$. According to the principle of reversibility, we also have $x_{ij} = x_{ji}$. 
The number of optical photons  $\vect{N}_{\rm pm}$ reaching all 64 photocathodes of the MAPMT  can therefore be given by
\begin{equation}
\vect{N}_{\rm pm}=\vecm{X} \vect{N}_{\rm bar}.
\label{eq:npm}
\end{equation}

The MAPMT transforms optical photons absorbed by its photocathode into electric signals on its anodes. 
These signals are subsequently read out by the FEE ASIC.
For gamma-ray energies detected by POLAR  one can assume that both the MAPMT and the FEE are in their linear range of responses.
Hence, the relationship between the recorded signal $\vect{E}_{\rm meas}$, i.e. the recorded energy depositions
and the number of the optical photons reaching MAPMT
$\vect{N}_{\rm pm}$
can be given by
\begin{equation}
\vect{E}_{\rm meas}= \vecm{G} \vect{N}_{\rm pm},
\label{eq:emeas}
\end{equation}
where $\vecm{G}={\rm Diag}\left( g_1, g_2, \cdots, g_{64}\right)$ 
represents the averaged signal induced by the optical photon in each of 64 channels.
It is worth stressing that $\vecm{G}$ is an individual feature of the MAPMT and FEE subsystem. 
In addition to the collection efficiency, $\vecm{G}$ also describes the module uniformity level.
Since both temperature and high voltage can change the system gain,
careful experimental calibrations are required to obtain correct value of $\vecm{G}$.
From Eqs. (\ref{eq:nbar}), (\ref{eq:npm}) and (\ref{eq:emeas}), we have
\begin{equation}
\vect{E}_{\rm meas}=(\vecm{G} \vecm{X} \vecm{B}) \vect{E}_{\rm dep} =\vecm{R} \vect{E}_{\rm dep},
\label{eq:resp}
\end{equation}
where $\vecm{R} =\left(r_{ij}\right)=\vecm{G} \vecm{X}\vecm{B}$ is
called the response matrix given by
\begin{equation}
r_{ij}=s g_i x_{ij} c_j.
\label{eq:rij}
\end{equation}
Since $x_{ij}$ represents the optical crosstalk between two channels, both $x_{ij}$ and $r_{ij}$
decrease with the distance between them.

The real energy $\vect{E}_{\rm dep}$ deposited by gamma-rays in all 64 PS bars can be reconstructed by a
linear transformation of the recorded energy depositions:
\begin{equation}
\vect{E}_{\rm dep}=\vecm{R}^\mathrm{-1} \vect{E}_{\rm meas}.
\label{eq:reconstruction}
\end{equation}
This transformation allows for a simultaneous correction of both crosstalk effects and non-uniformities. 
By knowing the values of both $\vect{E}_{\rm dep}$ and corresponding $\vect{E}_{\rm meas}$ one can also determine the response matrix $\vecm{R}$ .
The method used for its construction is described in the next chapter. 
It should be mentioned that this technique can be applied either  for laboratory tests on ground or during real observations in space.

\section{Determination of response matrix}

The energy deposited in PS bar in the process of Compton scattering is a function of the photon scattering angle.
It leads to a spectrum of energies with a sharp cutoff at the end of the spectrum called the Compton edge. 
The edge is related to the maximum energy transfered to the electron and full back-scattering of the gamma-ray.
It is commonly used for energy calibration in detector systems based on plastics scintillators such as POLAR.

Let us consider a scenario in which a gamma-ray is fully back-scattered in the $i$-th POLAR PS bar and 
escapes from the module. 
The energy is deposited in only one  $i$-th channel and is equal to the value of the Compton edge position 
$E_{\rm ce}=E_{\gamma}\left[1-1/(1+2E_{\gamma}/511)\right]$, where
$E_{\gamma}$ is the energy of the incident gamma-ray in units of keV. 
According to Eqs.~(\ref{eq:resp}) and ~(\ref{eq:rij}), the corresponding recorded
energy deposition
$E^{{\mathrm{ce}}}_i $, is given by
$ E^{\mathrm{ce}}_i
= r_{ii} \cdot E_{\rm ce}.$
As described in Ref.~\cite{silvio}, $E^{{\mathrm{ce}}}_i$ can be obtained by fitting the recorded energy spectrum with a smeared step-like function. 
Hence, the diagonal element $r_{ii}$ of the response matrix is given by:
\begin{equation}
r_{ii}=\frac{ E^{\mathrm{ce}}_i}{ E_{\rm ce}}.
\label{eq:riidet}
\end{equation}
The physical meaning of $ r_{ii}$ is the energy conversion factor of the $i$-th channel (in units of ADC channel/keV).
Applying the same procedure to all other channels provides the energy conversion matrix $\vecm{M}$ described as follows:
\begin{equation}
\begin{array}{c}
\vecm{M}=
\mathrm{Diag}\left(
r_{1,1}, r_{2,2}, \cdots, r_{64,64} \right).
\end{array}
\label{eq:ca}
\end{equation}
It should be noted that especially at lower gamma-ray energies the full energy absorption peaks from the photoelectric effect 
are frequently used for energy calibration of plastic scintillators.
In case of POLAR, calibrations with the Compton edge are easier than extracting of photo-peaks due to much lower
probability of the photoelectric process and poor PS energy resolution. In order to characterize POLAR response near its low energy threshold, 
several test measurements with the use of photopeaks were also conducted.

As described above, the diagonal elements of the response matrix can be determined by energy calibrations.
In order to determine the non-diagonal elements of the response matrix, let us again consider the scenario that 
the gamma-ray deposits energy only in the $i$-th channel. 
In such case, we have the corresponding
energy deposition vector $\vect{E}_{\rm dep}=( 0, \cdots, E^{\rm dep}_i , \cdots, 0)^{\rm T}$.
The recorded energy depositions of the $i$-th channel $E^{\rm meas}_i$ and
of the $j$-th channel $E^{\rm meas}_j$, according to Eqs.~(\ref{eq:resp}) and ~(\ref{eq:rij}),
are given by
\begin{equation}
E^{\rm meas}_i=r_{ii} E^{\rm dep}_i,
\end{equation}
and
\begin{equation}
E^{\rm meas}_j=r_{ji} E^{\rm dep}_i.
\label{eq:qj}
\end{equation}
Since the energy deposition only occurs only in the $i$-th channel, the  signal recorded in the $j$-th channel ($i \neq j$) comes from the crosstalk.
According to Eqs.~(\ref{eq:rij}) and ~(\ref{eq:qj}), the ratio $f_{ij}$ between the recorded energy depositions of the $i$-th and the $j$-th channel
, is called the crosstalk factor and is given by
\begin{equation}
f_{ij}=\frac{E^{\rm meas}_j}{E^{\rm meas}_i}=
\frac{r_{ji} \cdot E^{\rm dep}_i}{r_{ii} \cdot E^{\rm dep}_i}=\frac{r_{ji}}{r_{ii}}
=\frac{g_j x_{ji}}{g_i x_{ii}}.
\label{eq:fij}
\end{equation}
The crosstalk factor $f_{ij}$ represents partial transmission of the energy deposition from the $i$-th to the $j$-th channel. 
Note that $f_{ij}$ differs from the pure optical crosstalk factor $x_{ij}$ as it also includes  module non-uniformities.
Obviously, we have $f_{ij}=1$ for $i=j$, and $f_{ij}\neq f_{ji}$ for $i \neq j$ in most cases.
One way to obtain the crosstalk factor $f_{ij}$ is to determine the distribution of the recorded energy 
depositions of the $j$-th channel as the function of the recorded energy deposition of the $i$-th channel in Ref. \cite{silvio}.
For this purpose one can use pencil-like gamma-ray beams e.g. from 
the synchrotron light sources hitting exactly one PS bar. Either
Compton effect or photo-absorption process can be used for crosstalk analysis. 
POLAR instrument was characterized both ways using several
gamma-ray energies during test runs at the European Synchrotron Radiation Facility (ESRF) in Grenoble. 
It can be shown that the fit of the distribution with a line is usually sufficiently accurate for proper determination of the crosstalk factor.
Applying this procedure to all pairs of channels allows to construct the full 64 $\times$ 64 matrix $\vecm{F}=(f_{ij}) $ called the crosstalk matrix.
From Eqs.~(\ref{eq:resp}), (\ref{eq:rij}), (\ref{eq:ca}) and (\ref{eq:fij}) we find that the response matrix $\vecm{R}$ 
can be calculated from these two matrices using the equation:
\begin{equation}
\vecm{R} =\vecm{F}^{\rm T} \vecm{M}.
\label{eq:mres}
\end{equation}

\section{Model verification}

\begin{figure}[tb]
\centering
\includegraphics[width=0.6\textwidth]{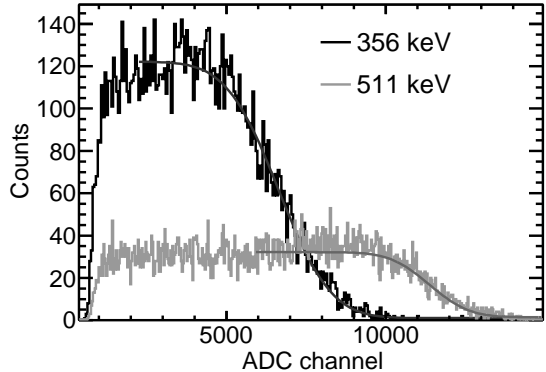}
\caption{Typical energy spectra from a single channel illuminated by X-ray beams with energies of 356
keV and 511 keV. The Compton edges provide energy conversion factors of 32.1 ADC channel/keV  and 33.4 ADC channel/keV respectively.
The difference is attributed to small non-linearities of the readout electronics.   
}
\label{fig:cefit}
\end{figure}

\begin{figure}[tb]
\centering
\includegraphics[width=0.8\textwidth]{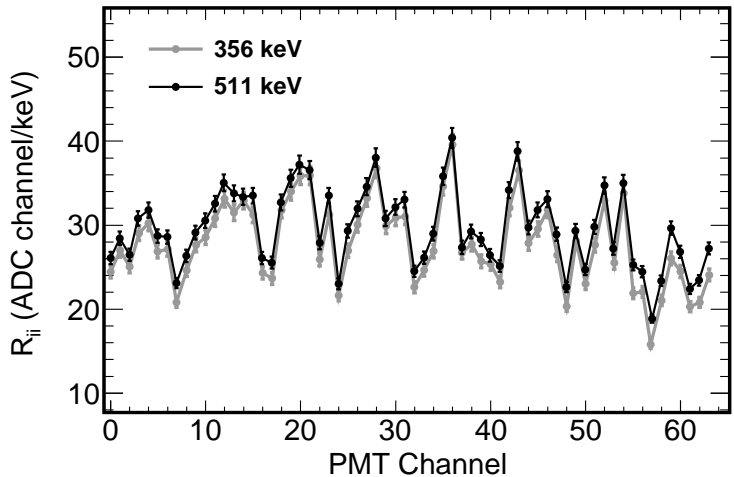}
\caption{Energy conversion factors (i.e. diagonal elements of the energy conversion matrix $\vecm{M}$) of one POLAR module 
obtained from the Compton edge fits of two energies.
The figure also indicates the response non-uniformity  within a module. 
}
\label{fig:cem}
\end{figure}

\begin{figure*}[htb]
\begin{center}
\includegraphics[width=0.8\textwidth]{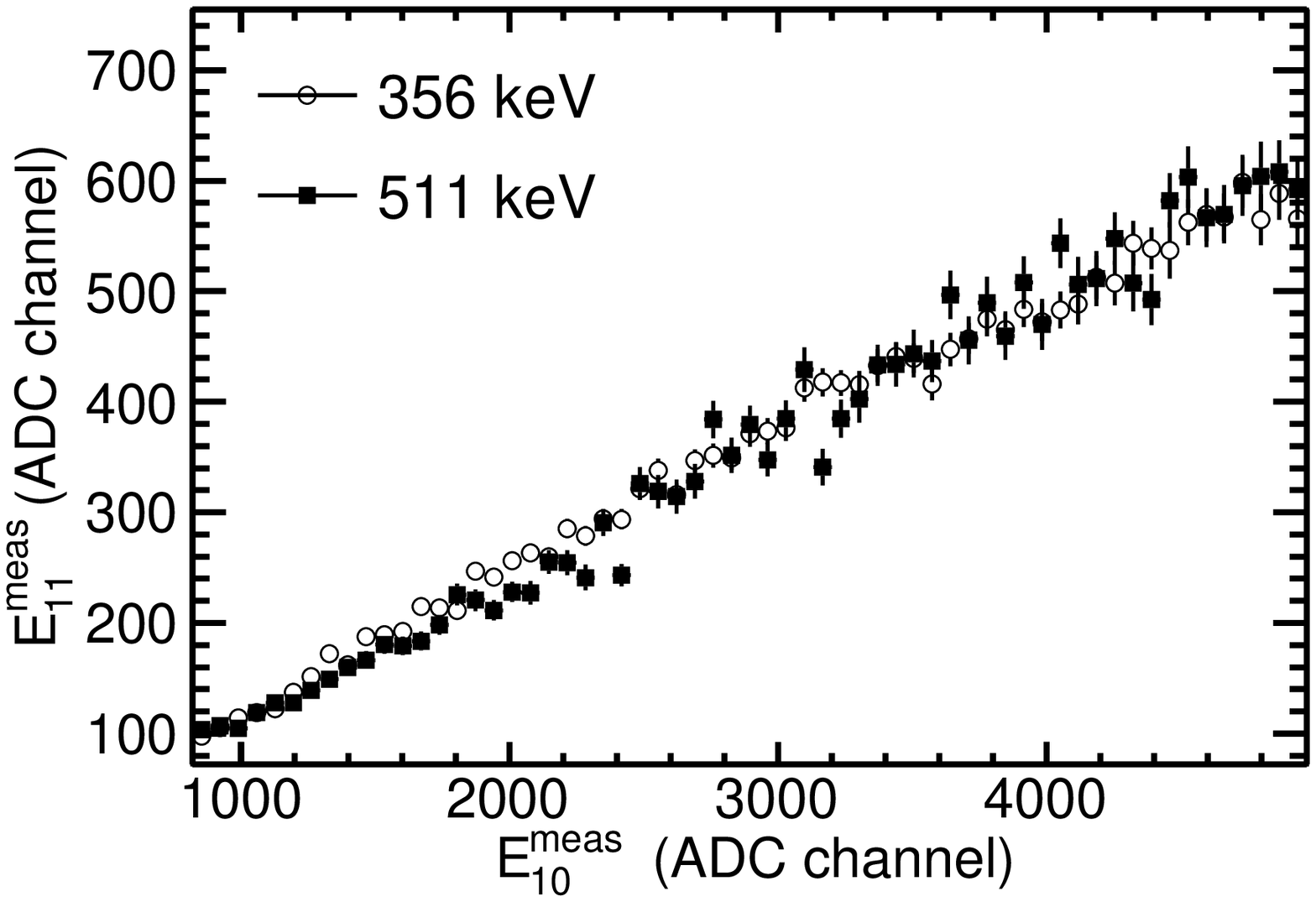}
\caption{
Crosstalk data as mean energy depositions in channel 11 as a function of the energy depositions in
channel 10 that was illuminated by X-ray beams with energies of 356 keV and 511 keV. 
Two crosstalk factors from the linear fits at two energies are equal to 12\%.
} 
\label{fig:chvsch}
\end{center}
\end{figure*}

We have conducted a series of extensive calibration runs of POLAR at ESRF with synchrotron
hard X-ray beams and at PSI using laboratory gamma-ray radioactive sources and lower energy X-rays from the X-ray generator.
Some of the low energy calibration runs and the studies of the photo-absorption peaks are described in Ref.~\cite{polarlowcal}. 
They were primary applied to determine responses near the low energy threshold and correct the ionization quenching effects that are 
significant at low energies \cite{zhangxf}.
For this analysis we use the data from calibration runs taken with the POLAR qualification model (QM) at the ESRF ID15A beamline
with the  X-ray energies of 356 keV and 511 keV. 
The synchrotron beam had  a square shape of 0.5$\times$0.5 mm$^2$. 
In order to avoid pile-ups the beam intensity was decreased to $\sim$ 2000 photons s$^{-1}$ using an Aluminum wedge. 
One module of the QM was mounted on the X-Y table driven by two step motors. 
They allowed to change the module position remotely and bring the beam to the center of each PS bar.
During the tests, each bar was illuminated for 10 seconds and then the next one was positioned at the beam.

As PS has better energy resolution at higher energies, both the 356 keV and the 511 keV X-rays were selected to determine the energy conversion factors. 
The procedure was used to verify and correct for possible response non-linearities e.g. in the electronic readout system.
Fig.~\ref{fig:cefit} shows typical energy spectra recorded by a single channel when its bar was illuminated with photon beams at energies of  511 keV 
and  356 keV.
The environmental background during the tests was negligible. 
The Compton edge position in each spectrum was found by fitting the right edge with the  step-like function:
\begin{equation}
f(x)=a_0+a_1\cdot {\rm Erfc}[(x-a_2)\cdot a_3],
\label{eq:efun1}
\end{equation}
where ${\rm Erfc}(x)$ is the error function given by
\begin{equation}
\mathrm{Erfc}(x)=\frac{2}{\sqrt{\pi}}\int_x^\infty \exp(-t^2) \mathrm{d}t.
\label{eq:efun2}
\end{equation}
The parameter $a_2$ is approximately equal to the position of the Compton edge in the units of ADC channel \cite{silvio}.
The energy conversion factor was calculated as $a_2/ E_{\rm ce}$, where $E_{\rm ce}$ is the Compton edge position in units of keV.
Fig.~\ref{fig:cem} shows the energy conversion factors for all 64 channels from the Compton edge fits.
The figure presents the  level of module non-uniformities and indicates some minor non-linearities in the energy response.
The non-linearity effect was due to increased walking of the triggers caused by  high values of the thresholds. It affected the 
instant of time of holding/sampling in the readout electronics. 
The mean value of two energy conversion factors
 was used to construct the energy conversion matrix $\vecm{M}$.

\begin{figure}[!hb]
\begin{center}
\includegraphics[width=0.7\textwidth]{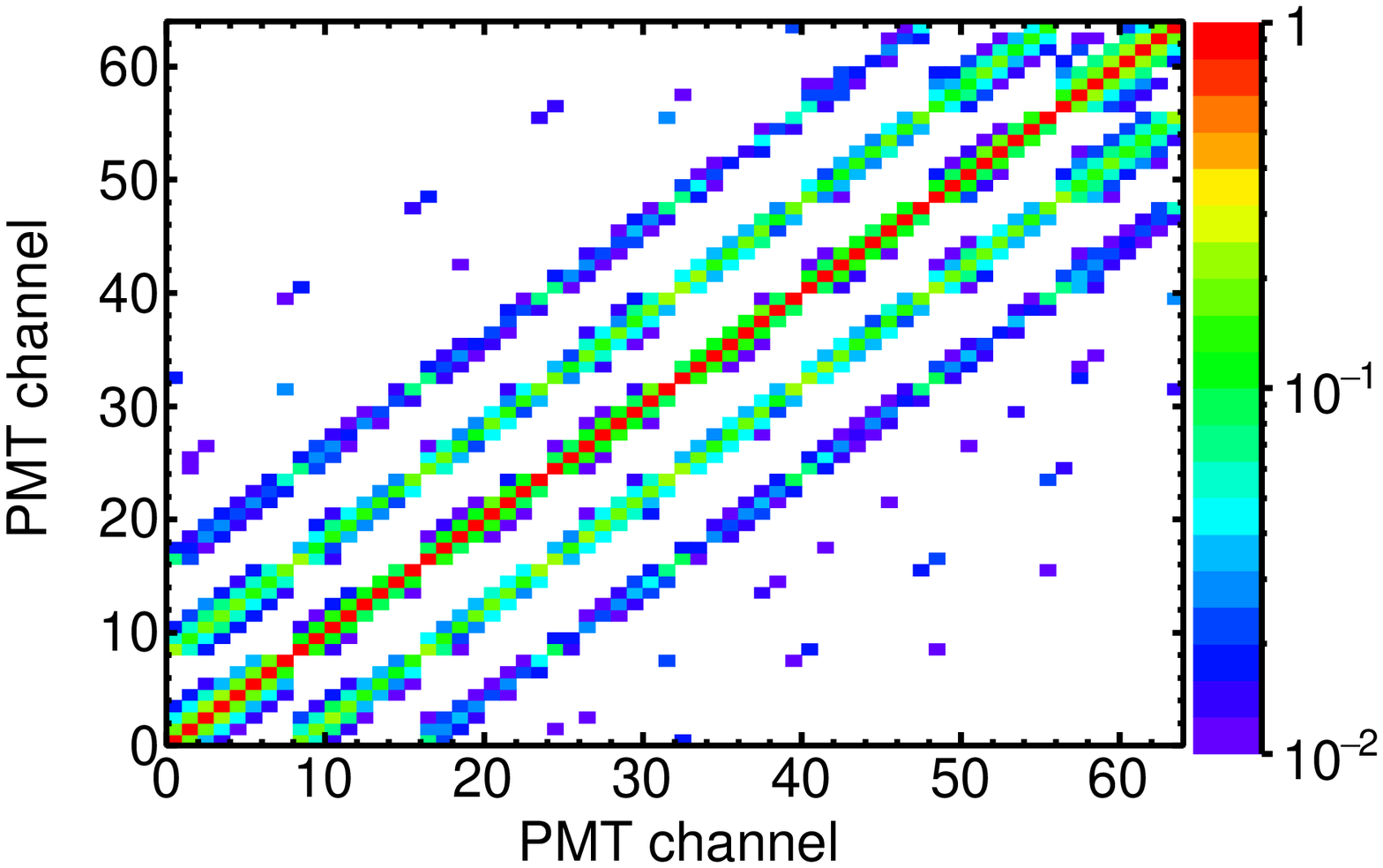}
\caption{An example of the crosstalk matrix $\vecm{F}$ of the selected module. 
Matrix elements represent crosstalk factors between two channels. 
Observed groups are due to the fact that the channel numbers between neighboring bars  
are not always continuous (see Fig.~\ref{fig:h8500}). The MAPMT channel numbering convention is shown in Fig.~\ref{fig:h8500}. 
}
\label{fig:crosstalkmatrix}
\end{center}
\end{figure}

\begin{figure}[!hb]
\begin{center}
\includegraphics[width=0.6\textwidth]{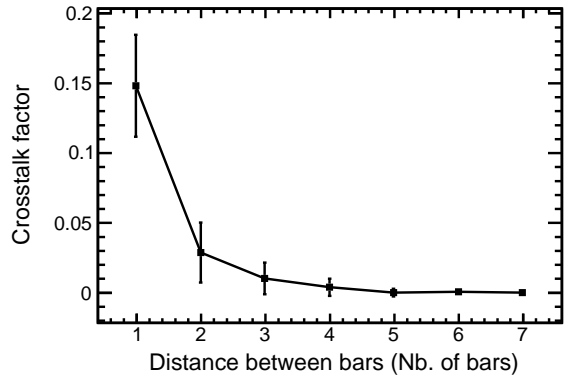}
\caption{
Mean crosstalk factor as a function of the distance between two bars calculated for all pairs within one module.
The mean value for the closest channels is on the level of 15\%.
}
\label{fig:crosstalkVdist}
\end{center}
\end{figure}

To determine the crosstalk we applied a method similar to one described in Ref.~\cite{silvio}.
In Fig.~\ref{fig:chvsch} we show an example of 
the mean energy depositions recorded by channel 11 as a function of energy deposition recorded by channel 10.
Only channel 10 was 
illuminated using either the 356 keV or the 511 keV X-ray beam.  Recorded energy 
depositions in all crosstalk channels are predominantly much smaller than depositions in the primary channel.
 Thus, only events for which the directly illuminated channel had larger recorded energy depositions were selected for data
 shown in Fig.~\ref{fig:chvsch}. 
According to the Monte Carlo simulations both the 511 keV and the 356 keV X-rays have less than 1\% chance to deposit energies
larger than 5 keV in one of its closest neighboring bars through direct scattering processes. 
Thus, the contribution of such scattered X-rays to the crosstalk distribution showed in Fig.~\ref{fig:chvsch} is negligible.
The data points can be fitted with a line and its  slope value represents the crosstalk factor. The slope values for the crosstalk 
from channel 10 to channel 11 are for both energies equal to 12\%.
It shows that it is sufficient to use just one X-ray energy to determine the crosstalk factor values.
Applying the same routine to correlation plots between any two channels of the same module provided the full crosstalk
matrix $\vecm{F}$ as shown in Fig.~\ref{fig:crosstalkmatrix}. 
The groups of channels observed in Fig.~\ref{fig:crosstalkmatrix} are due to the fact that the channel numbers of neighboring bars 
are not always continuous (see Fig.~\ref{fig:h8500}).
As shown in  Fig.~\ref{fig:crosstalkVdist}, the 
crosstalk factors strongly decrease with  increase of the distance between two channels. 
The mean crosstalk factor value between two closest neighboring channels is at the level of 15\%.
It should also be noted that the crosstalk factors can be determined from measurements with either
gamma-rays from radioactive sources or even an environmental background photons by applying  exactly the same method as above. 
For this purpose four weak $^{22}$Na radioactive sources were placed inside of the POLAR Flight Model (FM). 
As they emit two collinear annihilation photons the event selection will be unambiguous allowing for proper calibration of the whole instrument. 
It will be possible to  calibrate energies as well as determine crosstalk factors and non-uniformities during the entire flight onboard TG2.

\begin{figure}[!htb]
\centering
\includegraphics[width=0.6\textwidth]{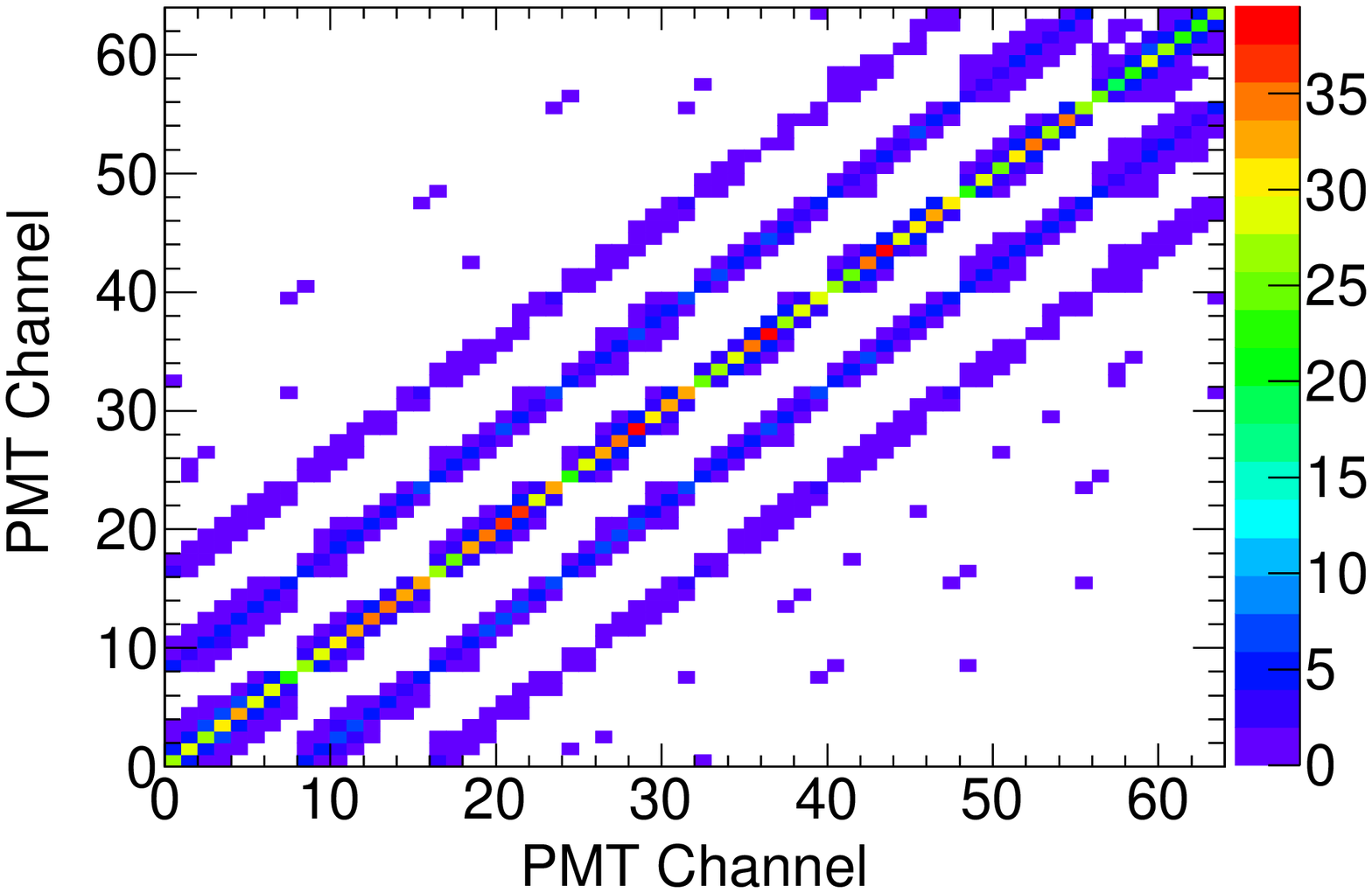}
\caption{Module response matrix $\vecm{R}$ calculated 
with Eq.~(\ref{eq:mres}). 
}
\label{fig:resM}
\end{figure}

\begin{figure*}[!htbp]
\begin{center}
\begin{minipage}{0.49\linewidth}
\includegraphics[width=1.1\textwidth]{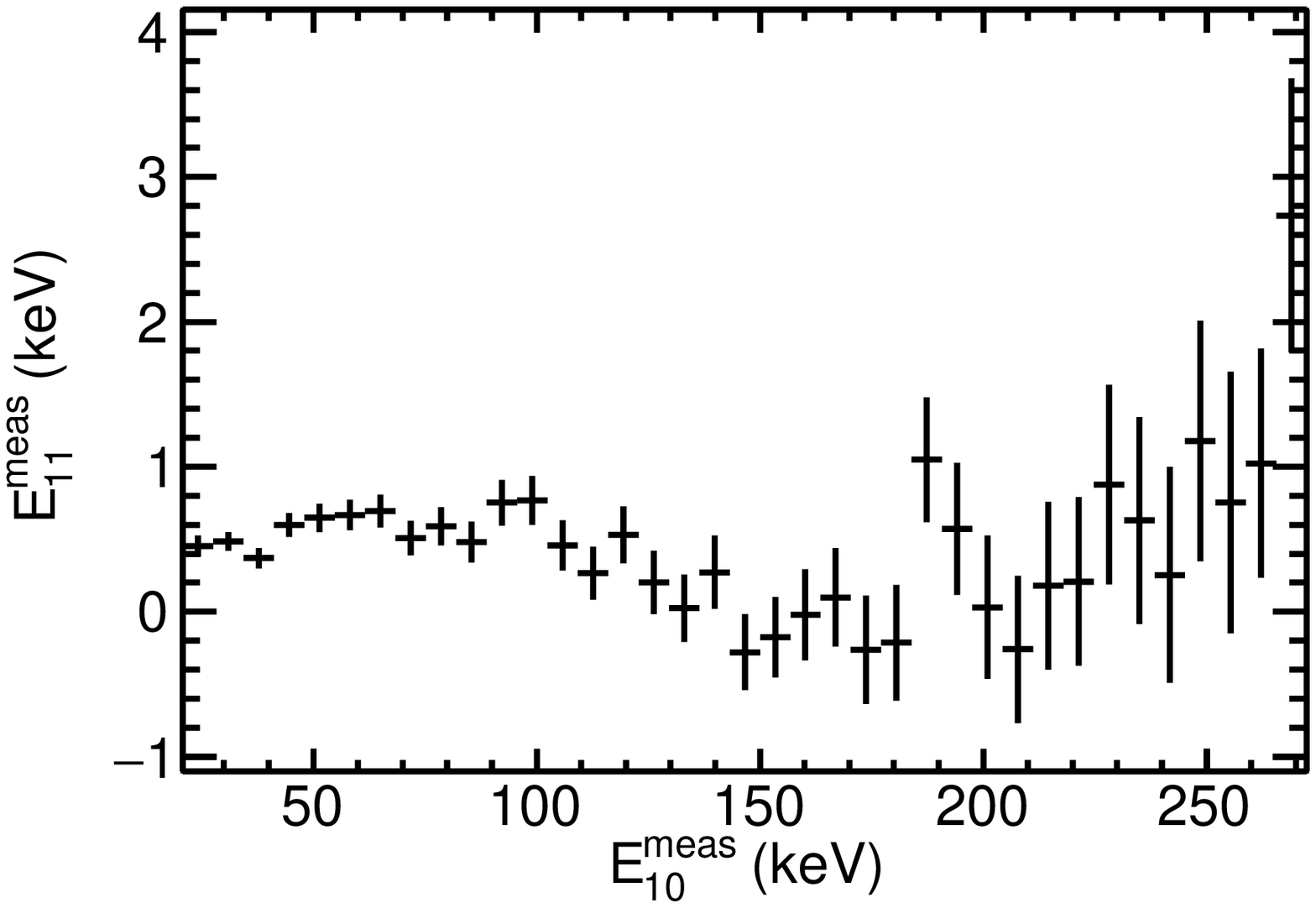}
\end{minipage}
\begin{minipage}{0.49\linewidth}
\includegraphics[width=1.1\textwidth]{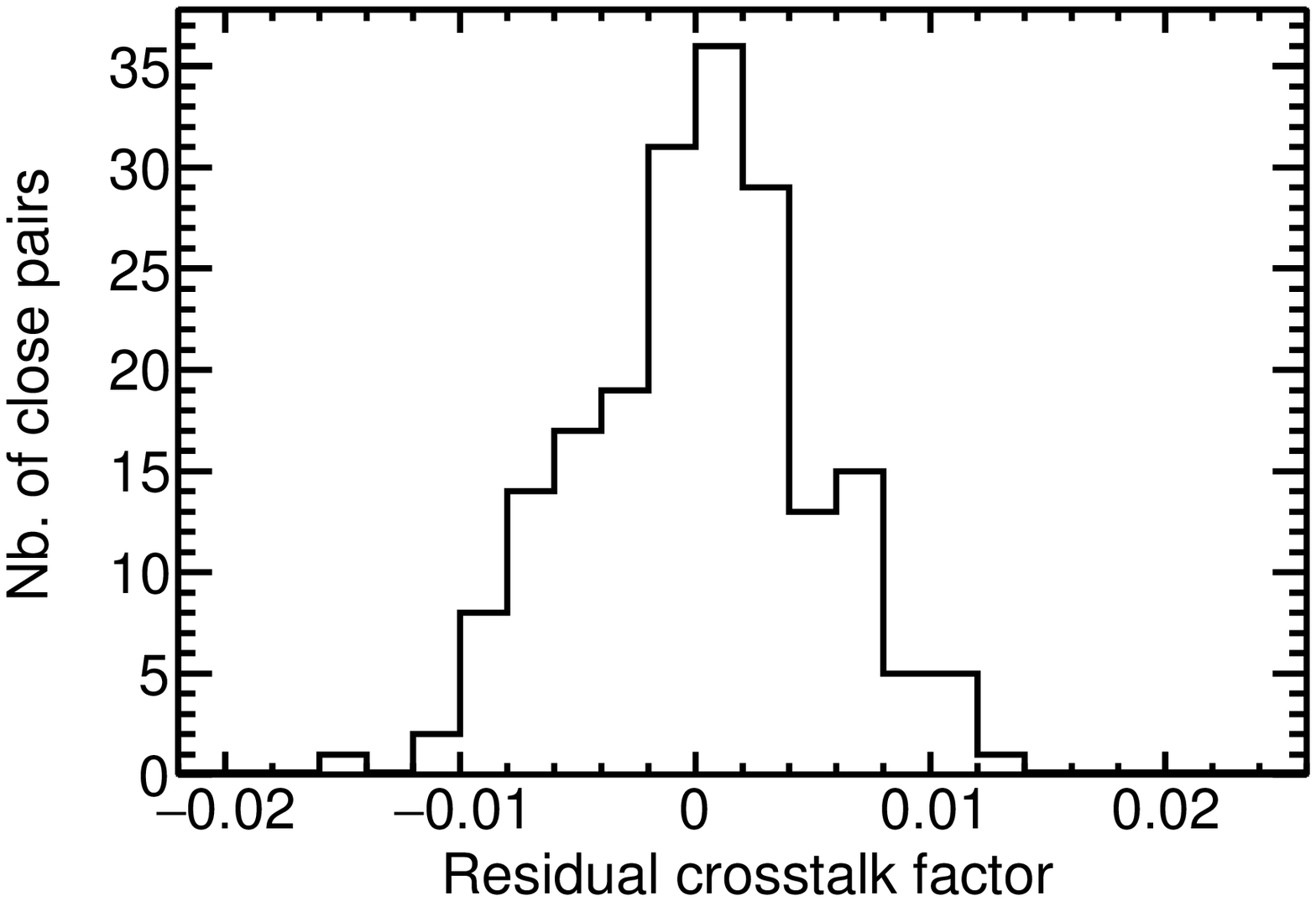}
\end{minipage}
\caption{
Residual crosstalk from channel 10 to channel 11 after applying corrections (left) and distribution of the residual crosstalk factors 
(i.e. the slopes of linear fits of the residual crosstalk plots) between two closest channels
(right).  The mean value of the residual crosstalk factors is equal to 0.01\% with a standard deviation of 0.5\%. 
The data from both runs at energies of  511 keV and 356 keV was used for the residual crosstalk studies.
}
\label{fig:residualxtalk}
\end{center}
\end{figure*}

Fig.~\ref{fig:resM} shows the response matrix $\vecm{R}$ calculated with Eq.~(\ref{eq:mres}).
Real energy depositions for two test runs were reconstructed with the help of $\vecm{R}$ according to Eq.~(\ref{eq:reconstruction}).
Both the crosstalk and the non-uniformities were to very large degree corrected.
As an example, the left panel of Fig.~\ref{fig:residualxtalk} shows the values of
the residual crosstalk from channel 10 to channel 11 in the corrected data.
We again fitted residual crosstalk plots with the first order polynomial 
to determine the remaing values.  
The distribution of  the residual crosstalk factors between two closest neighboring channels   
is shown in the right panel of Fig.~\ref{fig:residualxtalk}.
The mean value  is equal to 0.01\% with a standard deviation of 0.5\%.

\begin{figure}[H]
\begin{center}
\begin{minipage}{0.45\linewidth}
\includegraphics[width=1.1\textwidth]{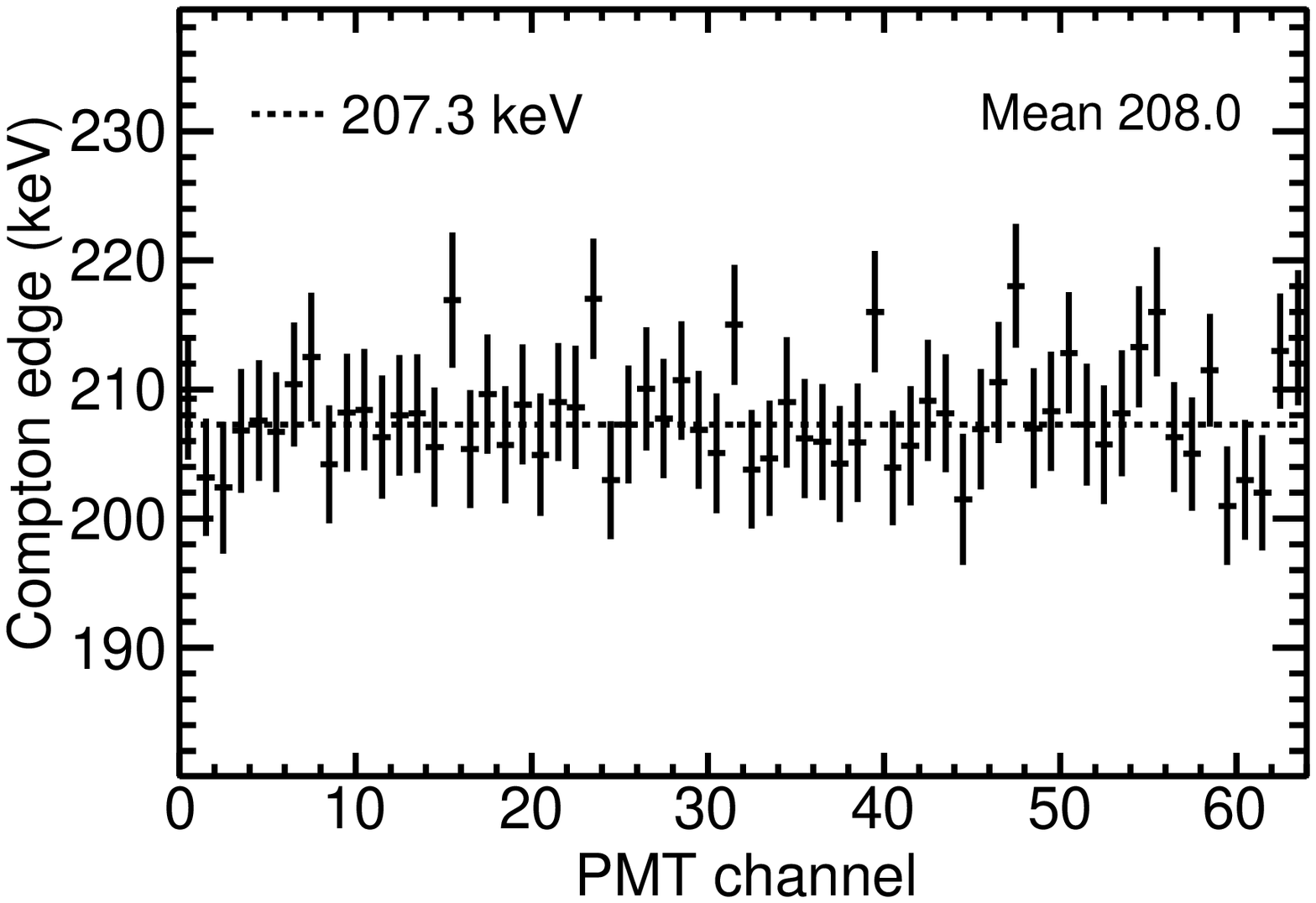}
\end{minipage}
\begin{minipage}{0.45\linewidth}
\includegraphics[width=1.1\textwidth]{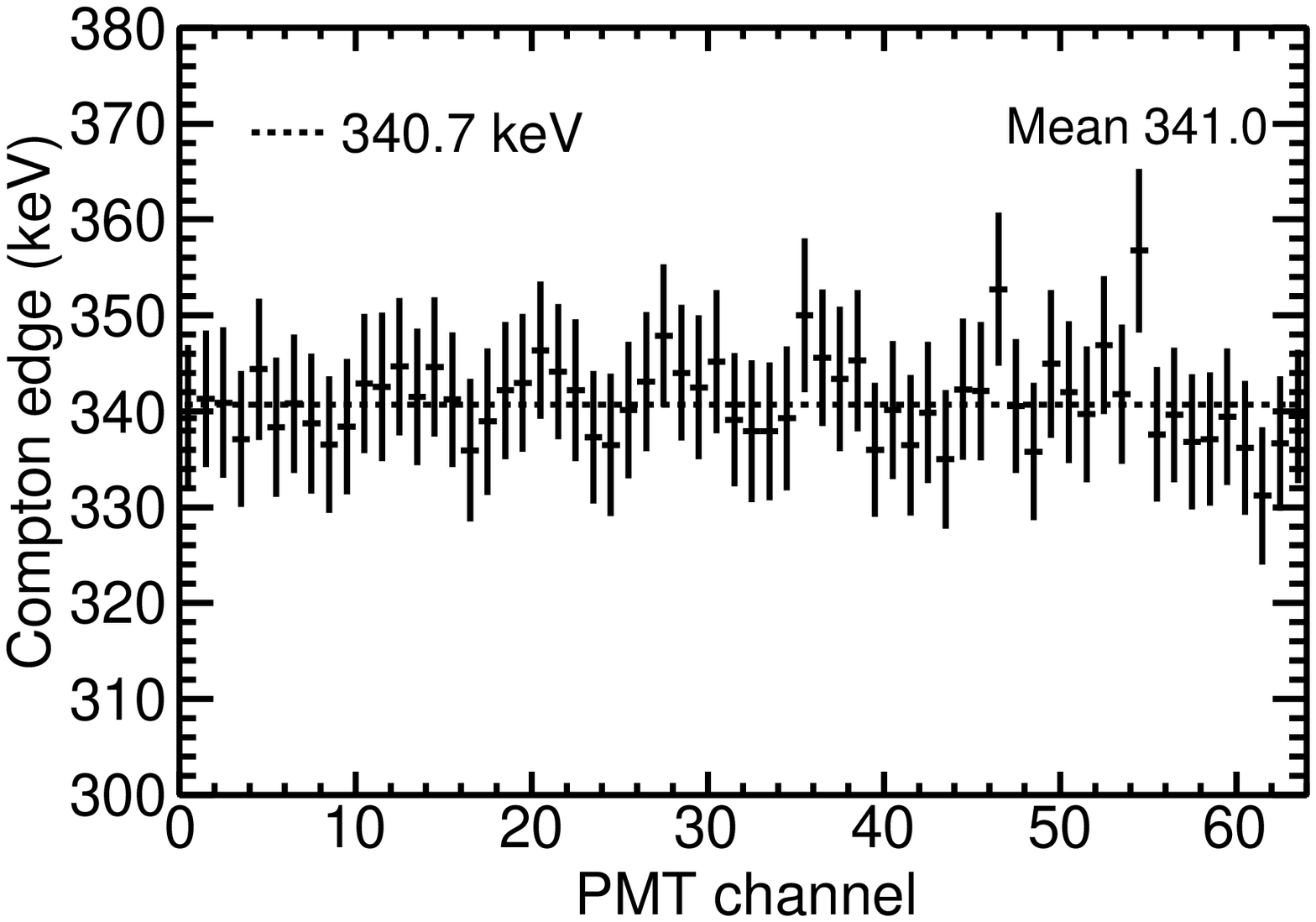}
\end{minipage}
\caption{Compton edge positions of the 64 channels in the reconstructed energy spectra for runs at energies of 356 keV (left) and 511 keV (right).
The crosstalk and non-uniformity corrections were applied to the selected events.  
Theoretical values of Compton edge energies are presented as dashed lines.
}
\label{fig:cecheck}
\end{center}
\end{figure}

In order to determine the correction of the module non-uniformities, we selected hits produced by the beams in the PS bars.
The same step-like function described in Eq.~(\ref{eq:efun1}) was used to fit the Compton edge positions in the energy spectra.
Compton edge positions of all 64 channels for runs at X-ray energies of 356 keV and 511 keV are shown in Fig.~\ref{fig:cecheck}.
The mean values of the corresponding Compton edge positions in the reconstructed energy spectra are 
208.0 keV and 341.0 keV, respectively. 
The theoretical values of Compton edge energies are equal to 207.3 keV and 340.7 keV respectively.
From the initial maximum non-uniformity ratio reaching the value of more than 200\% the final level found
after correcting the data is equal to 1.2\% for the 511 keV run and 2.0\% for the 356 keV run.

\section{Conclusion}

Polarization measurements in the new POLAR instrument are based on the Compton scattering and 
utilize distribution of the azimuthal angle for hard X-rays scattered between its 1600 plastic scintillator bars.
As the initial values of the POLAR response non-uniformity may reach more than a factor of two and the crosstalk factors could be higher than 20\% one must take these effects properly into account.
Therefore, for precise determination of the polarization observables in GRBs, a model describing the influence of the crosstalk and non-uniformities was developed.
Based on it, both factors can be corrected for in a rigorous way. The procedure uses a linear
transformation of the recorded energy depositions with a specially constructed response matrix.
A set of dedicated calibration runs was conducted for experimental determination and verification of the model parameters. 
The procedure involved two steps in which the energy calibration and crosstalk measurements were performed using mono-energetic, pencil-like X-ray beams.
The real energy depositions were obtained for each POLAR channel by applying the transformation formula as described above. 
To validate the method, the data from two test runs performed at ESRF using photons with energies of 356 keV and 511 keV were analysed.
Initial energy spectra were unfolded using previously constructed response matrix with implemented correction factors. 
The reconstructed data showed that the residual crosstalk factor between two neighboring channels was canceled almost 
to zero with a standard deviation of 0.5\%. 
The remaining non-uniformity values were at the level of less than 2\%.
The presented method provides a useful tool for more reliable computation of the real energies deposited in PS bars by the incoming photons.
It allows for precise determination of the polarization observables in the detected GRB events during POLAR operation in space. 
With some adjustments this method can also be used for other MAPMT-based instruments.

\section{Acknowledgments}
We gratefully acknowledge financial support from the National Basic Research Program (973 Program) of China under Grant No. 2014CB845800
 and the National Natural Science Foundation of China under Grant No. 11403028.

\bibliographystyle{model1-num-names}
\bibliography{theoryRef}

\begin{thebibliography}{17}
\expandafter\ifx\csname natexlab\endcsname\relax\def\natexlab#1{#1}\fi
\providecommand{\bibinfo}[2]{#2}
\ifx\xfnm\relax \def\xfnm[#1]{\unskip,\space#1}\fi
\bibitem[{Gomboc(2012)}]{grblasts}
\bibinfo{author}{A.~Gomboc},
\newblock \bibinfo{title}{Unveiling the secrets of gamma ray bursts},
\newblock \bibinfo{journal}{Contemporary Physics} \bibinfo{volume}{53}
  (\bibinfo{year}{2012}) \bibinfo{pages}{339--355}.
\bibitem[{Kumar and Zhang(2015)}]{zb3}
\bibinfo{author}{P.~Kumar}, \bibinfo{author}{B.~Zhang},
\newblock \bibinfo{title}{The physics of gamma-ray bursts \& relativistic
  jets},
\newblock \bibinfo{journal}{Physics Reports} \bibinfo{volume}{561}
  (\bibinfo{year}{2015}) \bibinfo{pages}{1 -- 109}. \bibinfo{note}{The physics
  of gamma-ray bursts \& relativistic jets}.
\bibitem[{Berger(2014)}]{review1}
\bibinfo{author}{E.~Berger},
\newblock \bibinfo{title}{Short-duration gamma-ray bursts},
\newblock \bibinfo{journal}{Annual Review of Astronomy and Astrophysics}
  \bibinfo{volume}{52} (\bibinfo{year}{2014}) \bibinfo{pages}{43--105}.
\bibitem[{Lazzati(2006)}]{zb2}
\bibinfo{author}{D.~Lazzati},
\newblock \bibinfo{title}{Polarization in the prompt emission of gamma-ray
  bursts and their afterglows},
\newblock \bibinfo{journal}{New Journal of Physics} \bibinfo{volume}{8}
  (\bibinfo{year}{2006}) \bibinfo{pages}{131}.
\bibitem[{Toma et~al.(2009)Toma, Sakamoto, Zhang, Hill, McConnell, Bloser,
  Yamazaki, Ioka, and Nakamura}]{zb1}
\bibinfo{author}{K.~Toma}, \bibinfo{author}{T.~Sakamoto},
  \bibinfo{author}{B.~Zhang}, \bibinfo{author}{J.~E. Hill},
  \bibinfo{author}{M.~L. McConnell}, \bibinfo{author}{P.~F. Bloser},
  \bibinfo{author}{R.~Yamazaki}, \bibinfo{author}{K.~Ioka},
  \bibinfo{author}{T.~Nakamura},
\newblock \bibinfo{title}{Statistical properties of gamma-ray burst
  polarization},
\newblock \bibinfo{journal}{The Astrophysical Journal} \bibinfo{volume}{698}
  (\bibinfo{year}{2009}) \bibinfo{pages}{1042}.
\bibitem[{Bloser et~al.(2009)Bloser, Legere, McConnell, Macri, Bancroft,
  Connor, and Ryan}]{grape}
\bibinfo{author}{P.~Bloser}, \bibinfo{author}{J.~Legere},
  \bibinfo{author}{M.~McConnell}, \bibinfo{author}{J.~Macri},
  \bibinfo{author}{C.~Bancroft}, \bibinfo{author}{T.~Connor},
  \bibinfo{author}{J.~Ryan},
\newblock \bibinfo{title}{Calibration of the gamma-ray polarimeter experiment
  (grape) at a polarized hard x-ray beam},
\newblock \bibinfo{journal}{Nuclear Instruments and Methods in Physics Research
  Section A: Accelerators, Spectrometers, Detectors and Associated Equipment}
  \bibinfo{volume}{600} (\bibinfo{year}{2009}) \bibinfo{pages}{424 -- 433}.
\bibitem[{Produit et~al.(2005)Produit, Barao, Deluit, Hajdas, Leluc, Pohl,
  Rapin, Vialle, Walter, and Wigger}]{nicolas}
\bibinfo{author}{N.~Produit}, \bibinfo{author}{F.~Barao},
  \bibinfo{author}{S.~Deluit}, \bibinfo{author}{W.~Hajdas},
  \bibinfo{author}{C.~Leluc}, \bibinfo{author}{M.~Pohl},
  \bibinfo{author}{D.~Rapin}, \bibinfo{author}{J.-P. Vialle},
  \bibinfo{author}{R.~Walter}, \bibinfo{author}{C.~Wigger},
\newblock \bibinfo{title}{Polar, a compact detector for gamma-ray bursts photon
  polarization measurements},
\newblock \bibinfo{journal}{Nuclear Instruments and Methods in Physics Research
  Section A: Accelerators, Spectrometers, Detectors and Associated Equipment}
  \bibinfo{volume}{550} (\bibinfo{year}{2005}) \bibinfo{pages}{616 -- 625}.
\bibitem[{Chauvin et~al.(2016)Chauvin, Flor{\'e}n, Jackson, Kamae, Kawano,
  Kiss, Kole, Mikhalev, Moretti, Olofsson, Rydstr{\"o}m, Takahashi, Lind,
  Str{\"o}mberg, Welin, Iyudin, Shifrin, and Pearce}]{pogolite}
\bibinfo{author}{M.~Chauvin}, \bibinfo{author}{H.-G. Flor{\'e}n},
  \bibinfo{author}{M.~Jackson}, \bibinfo{author}{T.~Kamae},
  \bibinfo{author}{T.~Kawano}, \bibinfo{author}{M.~Kiss},
  \bibinfo{author}{M.~Kole}, \bibinfo{author}{V.~Mikhalev},
  \bibinfo{author}{E.~Moretti}, \bibinfo{author}{G.~Olofsson},
  \bibinfo{author}{S.~Rydstr{\"o}m}, \bibinfo{author}{H.~Takahashi},
  \bibinfo{author}{J.~Lind}, \bibinfo{author}{J.-E. Str{\"o}mberg},
  \bibinfo{author}{O.~Welin}, \bibinfo{author}{A.~Iyudin},
  \bibinfo{author}{D.~Shifrin}, \bibinfo{author}{M.~Pearce},
\newblock \bibinfo{title}{The design and flight performance of the pogolite
  pathfinder balloon-borne hard x-ray polarimeter},
\newblock \bibinfo{journal}{Experimental Astronomy} \bibinfo{volume}{41}
  (\bibinfo{year}{2016}) \bibinfo{pages}{17--41}.
\bibitem[{Hill et~al.(2008)Hill, McConnell, Bloser, Legere, Macri, Ryan,
  Barthelmy, Angelini, Sakamoto, Black, Hartmann, Kaaret, Zhang, Ioka,
  Nakamura, Toma, Yamazaki, and Wu}]{poet}
\bibinfo{author}{J.~E. Hill}, \bibinfo{author}{M.~L. McConnell},
  \bibinfo{author}{P.~Bloser}, \bibinfo{author}{J.~Legere},
  \bibinfo{author}{J.~Macri}, \bibinfo{author}{J.~Ryan},
  \bibinfo{author}{S.~Barthelmy}, \bibinfo{author}{L.~Angelini},
  \bibinfo{author}{T.~Sakamoto}, \bibinfo{author}{J.~K. Black},
  \bibinfo{author}{D.~H. Hartmann}, \bibinfo{author}{P.~Kaaret},
  \bibinfo{author}{B.~Zhang}, \bibinfo{author}{K.~Ioka},
  \bibinfo{author}{T.~Nakamura}, \bibinfo{author}{K.~Toma},
  \bibinfo{author}{R.~Yamazaki}, \bibinfo{author}{X.~Wu},
\newblock \bibinfo{title}{Poet: Polarimeters for energetic transients},
\newblock \bibinfo{journal}{AIP Conference Proceedings} \bibinfo{volume}{1065}
  (\bibinfo{year}{2008}).
\bibitem[{Yonetoku et~al.(2011)Yonetoku, Murakami, Gunji, Mihara, Sakashita,
  Morihara, Kikuchi, Takahashi, Fujimoto, Toukairin, Kodama, Kubo, and
  Team}]{gap}
\bibinfo{author}{D.~Yonetoku}, \bibinfo{author}{T.~Murakami},
  \bibinfo{author}{S.~Gunji}, \bibinfo{author}{T.~Mihara},
  \bibinfo{author}{T.~Sakashita}, \bibinfo{author}{Y.~Morihara},
  \bibinfo{author}{Y.~Kikuchi}, \bibinfo{author}{T.~Takahashi},
  \bibinfo{author}{H.~Fujimoto}, \bibinfo{author}{N.~Toukairin},
  \bibinfo{author}{Y.~Kodama}, \bibinfo{author}{S.~Kubo},
  \bibinfo{author}{I.~D. Team},
\newblock \bibinfo{title}{Gamma-ray burst polarimeter (gap) aboard the small
  solar power sail demonstrator ikaros},
\newblock \bibinfo{journal}{Publications of the Astronomical Society of Japan}
  \bibinfo{volume}{63} (\bibinfo{year}{2011}) \bibinfo{pages}{625--638}.
\bibitem[{Hamamatsu(2011)}]{h8500}
\bibinfo{author}{Hamamatsu}, \bibinfo{title}{H8500 Datasheet},
  \bibinfo{year}{2011}.
\bibitem[{Muleri(2014)}]{polarconstruction}
\bibinfo{author}{F.~Muleri},
\newblock \bibinfo{title}{On the operation of x-ray polarimeters with a large
  field of view},
\newblock \bibinfo{journal}{The Astrophysical Journal} \bibinfo{volume}{782}
  (\bibinfo{year}{2014}) \bibinfo{pages}{28}.
\bibitem[{Orsi et~al.(2011)Orsi, Hajdas, Honkimäki, Lamanna, Lechanoine-Leluc,
  Marcinkowski, Pohl, Produit, Rapin, Suarez-Garcia, Rybka, and
  Vialle}]{silvio}
\bibinfo{author}{S.~Orsi}, \bibinfo{author}{W.~Hajdas},
  \bibinfo{author}{V.~Honkimäki}, \bibinfo{author}{G.~Lamanna},
  \bibinfo{author}{C.~Lechanoine-Leluc}, \bibinfo{author}{R.~Marcinkowski},
  \bibinfo{author}{M.~Pohl}, \bibinfo{author}{N.~Produit},
  \bibinfo{author}{D.~Rapin}, \bibinfo{author}{E.~Suarez-Garcia},
  \bibinfo{author}{D.~Rybka}, \bibinfo{author}{J.-P. Vialle},
\newblock \bibinfo{title}{Response of the compton polarimeter polar to
  polarized hard x-rays},
\newblock \bibinfo{journal}{Nuclear Instruments and Methods in Physics Research
  Section A: Accelerators, Spectrometers, Detectors and Associated Equipment}
  \bibinfo{volume}{648} (\bibinfo{year}{2011}) \bibinfo{pages}{139 -- 154}.
\bibitem[{Montgomery et~al.(2015)Montgomery, Hoek, Lucherini, Mirazita,
  Orlandi, Pereira, Pisano, Rossi, Viticchiè, and Witchger}]{mapmtRef2}
\bibinfo{author}{R.~Montgomery}, \bibinfo{author}{M.~Hoek},
  \bibinfo{author}{V.~Lucherini}, \bibinfo{author}{M.~Mirazita},
  \bibinfo{author}{A.~Orlandi}, \bibinfo{author}{S.~A. Pereira},
  \bibinfo{author}{S.~Pisano}, \bibinfo{author}{P.~Rossi},
  \bibinfo{author}{A.~Viticchiè}, \bibinfo{author}{A.~Witchger},
\newblock \bibinfo{title}{Investigation of hamamatsu \{H8500\} phototubes as
  single photon detectors},
\newblock \bibinfo{journal}{Nuclear Instruments and Methods in Physics Research
  Section A: Accelerators, Spectrometers, Detectors and Associated Equipment}
  \bibinfo{volume}{790} (\bibinfo{year}{2015}) \bibinfo{pages}{28 -- 41}.
\bibitem[{Sun et~al.(2011)Sun, Wu, Zhang, Lu, Li, Dong, Li, Chai, Zhang, Kang,
  and Song}]{sunjc}
\bibinfo{author}{J.~Sun}, \bibinfo{author}{B.~Wu}, \bibinfo{author}{Y.~Zhang},
  \bibinfo{author}{Y.~Lu}, \bibinfo{author}{Y.~Li}, \bibinfo{author}{Y.~Dong},
  \bibinfo{author}{L.~Li}, \bibinfo{author}{J.~Chai},
  \bibinfo{author}{Y.~Zhang}, \bibinfo{author}{S.~Kang},
  \bibinfo{author}{L.~Song},
\newblock \bibinfo{title}{A prototype study of the polar front-end
  electronics},
\newblock \bibinfo{journal}{Nuclear Instruments and Methods in Physics Research
  Section A: Accelerators, Spectrometers, Detectors and Associated Equipment}
  \bibinfo{volume}{659} (\bibinfo{year}{2011}) \bibinfo{pages}{322 -- 327}.
\bibitem[{Xiao et~al.(2015)Xiao, Hajdas, Bao, Batsch, Bernasconi, Cernuda,
  Chai, Dong, Gauvin, Kole, Kong, Kong, Li, Liu, Liu, Marcinkowski, Orsi, Pohl,
  Produit, Rapin, Rutczynska, Rybka, Shi, Song, Sun, Szabelski, Wu, Wang, Wen,
  Xu, Zhang, Zhang, Zhang, Zhang, Zhang, and Zwolinska}]{polarlowcal}
\bibinfo{author}{H.~L. Xiao}, \bibinfo{author}{W.~Hajdas},
  \bibinfo{author}{T.~W. Bao}, \bibinfo{author}{T.~Batsch},
  \bibinfo{author}{T.~Bernasconi}, \bibinfo{author}{I.~Cernuda},
  \bibinfo{author}{J.~Y. Chai}, \bibinfo{author}{Y.~W. Dong},
  \bibinfo{author}{N.~Gauvin}, \bibinfo{author}{M.~Kole},
  \bibinfo{author}{M.~N. Kong}, \bibinfo{author}{S.~W. Kong},
  \bibinfo{author}{L.~Li}, \bibinfo{author}{J.~T. Liu},
  \bibinfo{author}{X.~Liu}, \bibinfo{author}{R.~Marcinkowski},
  \bibinfo{author}{S.~Orsi}, \bibinfo{author}{M.~Pohl},
  \bibinfo{author}{N.~Produit}, \bibinfo{author}{D.~Rapin},
  \bibinfo{author}{A.~Rutczynska}, \bibinfo{author}{D.~Rybka},
  \bibinfo{author}{H.~L. Shi}, \bibinfo{author}{L.~M. Song},
  \bibinfo{author}{J.~C. Sun}, \bibinfo{author}{J.~Szabelski},
  \bibinfo{author}{B.~B. Wu}, \bibinfo{author}{R.~J. Wang},
  \bibinfo{author}{X.~Wen}, \bibinfo{author}{H.~H. Xu},
  \bibinfo{author}{L.~Zhang}, \bibinfo{author}{L.~Y. Zhang},
  \bibinfo{author}{S.~N. Zhang}, \bibinfo{author}{X.~F. Zhang},
  \bibinfo{author}{Y.~J. Zhang}, \bibinfo{author}{A.~Zwolinska},
  \bibinfo{title}{Calibration of the gamma-ray burst polarimeter polar},
  \bibinfo{year}{2015}.
\bibitem[{Zhang et~al.(2015)Zhang, Xiao, Yu, Orsi, Wu, Hu, and Zhang}]{zhangxf}
\bibinfo{author}{X.~Zhang}, \bibinfo{author}{H.~Xiao}, \bibinfo{author}{B.~Yu},
  \bibinfo{author}{S.~Orsi}, \bibinfo{author}{B.~Wu}, \bibinfo{author}{W.~Hu},
  \bibinfo{author}{X.~Zhang},
\newblock \bibinfo{title}{Study of non-linear energy response of \{POLAR\}
  plastic scintillators to electrons},
\newblock \bibinfo{journal}{Nuclear Instruments and Methods in Physics Research
  Section A: Accelerators, Spectrometers, Detectors and Associated Equipment}
  \bibinfo{volume}{797} (\bibinfo{year}{2015}) \bibinfo{pages}{94 -- 100}.

\end{thebibliography}
\end{document}